\documentclass[a4paper,12pt]{article}
\usepackage{graphicx}
\usepackage[english]{babel}
\usepackage{amssymb,amsbsy,amsmath,eucal}
\usepackage[compress]{cite}
\newcommand{\ha}{\mbox{\small$\frac{1}{2}$}}
\newcommand{\lab}[1]{\label{#1}}
\newcommand{\re}[1]{(\ref{#1})}
\newcommand{\nn}{\nonumber}
\newcommand{\B}[1]{\boldsymbol{#1}}

\newcommand{\s}[1]{\mathsf{#1}}

\newcommand{\sOm}{\mathsf{\Omega}}
\newcommand{\BOm}{\boldsymbol{\Omega}}

\newcommand{\D}[2]{{\rm d}^{#1}{#2}\,}
\begin{document}

\title{Rotary dynamics of the rigid body electric dipole under the radiation reaction}

\author{A. Duviryak}
\date{Institute for Condensed Matter Physics of NAS of Ukraine,\\
1 Svientsitskii Street, Lviv, UA-79011, Ukraine \\
              Tel.: +380 322 701496, \
              Fax: +380 322 761158\\
              {duviryak@icmp.lviv.ua}           
}

\maketitle

\begin{abstract}
Rotation of a permanently polarized rigid body under the radiation
reaction torque is considered. Dynamics of the spinning top is derived from a balance
condition of the angular momentum. It leads to the non-integrable
nonlinear 2nd-order equations for angular velocities, and then to
the reduced 1st-order Euler equations. The example of an axially
symmetric top with the longitudinal dipole is solved exactly, with
the transverse dipole is analyzed qualitatively and numerically.
Physical solutions describe the asymptotic power-law slowdown to
stop or the exponential drift to a residual rotation;
this depends on initial conditions and a shape of the top. \\
Keywords: Radiation reaction, spinning top\\
PACS: 41.60.-m, 45.40.-f, 02.30.Ik
\end{abstract}


\section{Introduction}
\renewcommand{\theequation}{1.\arabic{equation}}
\setcounter{equation}{0}

It was reported recently that silica particles of mass about 1\,fg
and size 100\,nm were spined up in the optical trap to the frequency
above 1\,GHz \cite{RDHD18}\footnote{This record has been improved
to 5.2\,GHz\cite{AXBJGL20}.}. This corresponds to the orbital quantum
number of order $\ell\gtrsim10^{10}$, i.e., the rotational motion is
quite classical. If such a particle possesses an electric dipole
moment then it emits the electromagnetic radiation and receives the
reaction torque which slows the particle rotation down. Of course,
this relativistic effect is very small for the aforementioned
particle whose constituents move no faster than 10$^{-6}$ of light
speed. But if the spinning particle is kept free for a long time,
its radiative slowdown can be measurable and deserves a
theoretical study.

The classical dynamics of a single point-like charge is governed by
the relativistic {\em Lorentz-Dirac} equation or its slow-motion
predecessor, the {\em Abraham-Lorentz} equation \cite{Roh90,YaT12}.
Both include the radiation reaction terms which depend on 3rd-order
derivatives and give rise to redundant runaway solutions. To get rid
of these solutions one uses frequently the approximated 2nd-order
reduction of the Abraham-Lorentz-Dirac equations in which higher
derivatives in r.-h.s. are eliminated by means of the same but
truncated equation (i.e., without a radiation reaction term)
\cite{L-L87E,Spo00,Par06}.

For a composite spinning particle the dynamics is complemented with
rotary degrees of freedom and complicated by further electrodynamical effects,
such as an interaction between different charges, interference of
radiation outgoing from them etc. As a result,
the external and  radiation reaction forces are weaved in a complex way
together in a complete set of equations on motion \cite{dGS72}
whose cumbersome form leaves little feasibility for their analysis.

In the present paper we consider a free non-relativistic composite
particle with an electric dipole moment. In this case the
translational motion is inertial while of interest is a nontrivial
3D-rotary dynamics under the radiation reaction torque. We utilize the
rigid body kinematics and the angular momentum balance condition in
slow-motion approximation \cite[\S75]{L-L87E}, and derive the
Euler-type equations of motion. Two specific examples of
axially-symmetric top are considered: with longitudinal and
transverse dipole moments. The first example is solvable, the second
one is not, thus it is analyzed qualitatively and numerically. Our
methodology is illustrated by the simple case of a plane rotator.


\section{Dynamics of a single charge plane rotator}
\renewcommand{\theequation}{2.\arabic{equation}}
\setcounter{equation}{0}

The slowdown of a charged plane rotator can be deduced easily from the
energy balance condition, by means of the Larmor formula
\cite{L-L87E,Roh90,YaT12}. Instead, we proceed by methodological
purpose from the Abraham-Lorentz equation:
%
\begin{equation}\lab{2.1}
m\dot{\B v}=\B F+\frac{2q^2}{3c^3}\ddot{\B v},
\end{equation}
where $\B v$ and $\dot{\B v}\equiv\D{}{\B v}/\D{}{t}$ are velocity and acceleration
of particle of mass $m$ and charge $q$, and $c$ is the speed of light.
The last term in r.-h.s. of eq. \re{2.1} describes the radiation reaction force;
it causes physical effect only in the presence of other forces $\B F$.

Let $\B F$ be a reaction of holonomic constraint which admits a
circular motion, i.e., turns the particle into the plane rotator.
Then the equation \re{2.1} reduces to the form:
%
\begin{equation}\lab{2.2}
\dot\Omega=\tau_0(\ddot\Omega-\Omega^3),\qquad\quad\mbox{where}\qquad \tau_0=\frac{2q^2}{3mc^3},
\end{equation}
which henceforth will be referred to as the {\em rotator Lotentz}
equation. This is the 2nd-order equation for the angular velocity
$\Omega$. Moreover, similarly to the Abraham-Lorentz equation
\re{2.1}, equation \re{2.2} is singularly perturbed, i.e., there
is a small time-scale parameter $\tau_0$  in r.-h.s. at the
higher-order derivative $\ddot\Omega$. Thus this equation possesses
redundant set of solutions which must be separated out from the
physical solutions.

Exact solutions of \re{2.2} are unknown. In Appendix A the equation \re{2.2} is analyzed
qualitatively and numerically, and selection rules for the physical solutions are formulated.

Here we use in the r.-h.s. of the equation \re{2.2} the truncated form of this equation,
$\dot\Omega=O(\tau_0)$, and arrive at the approximately reduced 1st-order equation:
%
\begin{equation}\lab{2.3}
\dot\Omega=-\tau_0\Omega^3.
\end{equation}
It admits a standard Cauchy problem and possesses the following
solution:

%
\begin{equation}\lab{2.4}
\Omega(t)=\frac{\Omega_0}{\sqrt{1+2\tau_0\Omega_0^2t}}, \quad\quad
\Omega(0)=\Omega_0.
\end{equation}
The characteristic quantity $T=1/(\tau_0\Omega_0^2)$ is a time
during which a rotary braking is most intense. Asymptotically, at $t\gg T$,
the angular velocity decreases by the power law
$\Omega\sim1/\sqrt{2\tau_0t}$ which does not depend on the initial
value $\Omega_0$. In Appendix A this result is obtained by the
asymptotical analysis of the rotator Lorentz equation \re{2.2}.


\section{Equation of motion of a polarized spinning top}
\renewcommand{\theequation}{3.\arabic{equation}}
\setcounter{equation}{0}

A plane rotator is the simplest spinning system.
Here we consider a composite particle consisting of
point-like charges $q_a$ ($a~=~1,2,\dots)$ with masses $m_a$
located at positions $\B r_a$. We proceed from the slow-motion balance equation \cite[\S75]{L-L87E}:
%
\begin{equation}\lab{3.1}
\dot{\B L}=\frac2{3c^3}\,\B{\frak d}\times\dddot{\B{\frak d}}
\end{equation}
for the angular momentum $\B L=\sum_a m_a\,\B r_a\times\B v_a$, where
$\B{\frak d}=\sum_a q_a\B r_a$ is the dipole moment of the system.

The composite particle is considered as a free rigid body, i.e., a
top. A rotational motion of an arbitrary point $\B r(t)$ of the top
can be presented as follows: $\B r(t)=\s O(t)\B\rho$, where $\s
O(t)\in\ $SO(3) is a rotation matrix, and $\B\rho$ is a constant
position of the point in the proper reference frame of
the top. We will use the following kinematic relations:
%
\begin{eqnarray}\lab{3.2}
\B v&\equiv&\dot{\B r}=\dot{\s O}\B\rho=\s O(\BOm\times\B\rho),\nn\\
\dot{\B v}&=&\s O\{\BOm\times(\BOm\times\B\rho)+\dot\BOm\times\B\rho\},\nn\\
\ddot{\B v}&=&\s
O\{\BOm\times[\BOm\times(\BOm\times\B\rho)+2\dot\BOm\B\rho]+\dot\BOm\times(\BOm\times\B\rho)+\ddot\BOm\times\B\rho\},
\end{eqnarray}
where $\BOm$ is the angular velocity vector, dual to the skew matrix
$\sOm\equiv\s O^{\rm T}\dot{\s O}$.

Using the relations \re{3.2} in eq. \re{3.1} yields the equation of rotary motion of a top:
%
\begin{eqnarray}\lab{3.3}
\s I\dot\BOm+\BOm\times\s I\BOm
&=&\frac2{3c^3}\B d\times\{\B d\times(\Omega^2\BOm-\ddot\BOm)+(\B d\cdot\dot\BOm)\BOm+2(\B d\cdot\BOm)\dot\BOm\};\qquad
\end{eqnarray}
here $\s I=||I_{ij}||$ ($i,j=1,2,3$) is the inertia tensor, $\B d\equiv\s O^{\rm T}\B{\frak d}=\sum_a m_a\B\rho_a$
is a constant dipole moment of the top in the proper reference frame, and $\Omega\equiv|\BOm|$.

The 1st-order reduction of the equation \re{3.3} implies
the elimination the 2nd-order derivative in r.-h.s.
by means of the truncated equation and its differential consequence:
%
\begin{eqnarray}\lab{3.4}
\dot\BOm&=&-\s I^{-1}(\BOm\times\s I\BOm),\nn\\
\ddot\BOm&=&-\s I^{-1}(\dot\BOm\times\s I\BOm+\BOm\times\s I\dot\BOm)\nn\\
&=&\s I^{-1}\{(\BOm\times(\BOm\times\s I\BOm)-(\s I\BOm)\times\s I^{-1}(\BOm\times\s I\BOm)\}.
\end{eqnarray}
General explicit form of the reduced equation is cumbersome and omitted here.

\subsection{Dynamics of an axially symmetric spinning top with the longitudinal dipole}

Here we limit ourself by the case of spinning top with an axially-symmetric inertia ellipsoid (i.e., spheroid) and the longitudinal dipole moment:
%
\begin{equation}\lab{3.5}
I_{ij}=I_i\delta_{ij}, \quad I_2=I_1;\qquad d_1=d_2=0,\quad
d_3\equiv d.
\end{equation}
Then the equation \re{3.3} splits into the following set:
%
\begin{eqnarray}
I_1\dot\Omega_1&=&(I_1-I_3)\Omega_2\Omega_3+\frac{2d^2}{3c^3}\left\{\ddot\Omega_1-\Omega^2\Omega_1-\dot\Omega_3\Omega_2-2\Omega_3\dot\Omega_2\right\},
\lab{3.6}\\
I_1\dot\Omega_2&=&(I_3-I_1)\Omega_1\Omega_3+\frac{2d^2}{3c^3}\left\{\ddot\Omega_2-\Omega^2\Omega_2+\dot\Omega_3\Omega_1+2\Omega_3\dot\Omega_1\right\},
\lab{3.7}\\
I_3\dot\Omega_3&=&0.
\lab{3.8}
\end{eqnarray}

It follows from eq. \re{3.8} that $\Omega_3=\,$const provided
$I_3\ne0$. Otherwise $\Omega_3$ remains an undetermined function of time.
From the physical viewpoint, in the case $I_3\to0$
a rigid body degenerates to an infinitely thin rod along
the $O\rho_3$ axis, and a rotation in this direction is indefinite.

The remaining set of equations \re{3.6}--\re{3.7} is not solved exactly.
In Appendix B an asymptotic behavior of their solutions
at $t\to\pm\infty,0$ is analyzed.

Here we consider the 1st-order reduction of these equations:
%
\begin{eqnarray}
I_1\dot\Omega_1&=&(I_1-I_3)\Omega_2\Omega_3-\frac{2d^2}{3c^3}\left\{\Omega_1^2+\Omega_2^2+\frac{I_3^2}{I_1^2}\Omega_3^2\right\}\Omega_1,
\lab{3.9}\\
I_1\dot\Omega_2&=&(I_3-I_1)\Omega_1\Omega_3-\frac{2d^2}{3c^3}\left\{\Omega_1^2+\Omega_2^2+\frac{I_3^2}{I_1^2}\Omega_3^2\right\}\Omega_2.
\lab{3.10}
\end{eqnarray}

If $I_3\ne0$, the equation \re{3.8} yields $\Omega_3=\,$const.
The remaining nonlinear equations \re{3.9}--\re{3.10} determine
components $\Omega_1$, $\Omega_2$ which form the vector $\BOm_\bot=\{\Omega_1,\Omega_2,0\}$.
Then one multiplies the equation \re{3.9} by $\Omega_1$, the equation \re{3.10} by $\Omega_2$,
so their sum yields an integrable equation for $\Omega_\bot^2\equiv\BOm_\bot^2$.
Substituting this integral back into the set \re{3.9}--\re{3.10} reduces
the latter to a linear set. A final integration yields the solution:
%
\begin{eqnarray}
&&\Omega_1=\Omega_\bot\,\cos{\tilde\Omega_3t},\quad\Omega_2=-\Omega_\bot\,\sin{\tilde\Omega_3t},
\lab{3.11}\\
&& \Omega_\bot\equiv|\BOm_\bot|=
\frac{(I_3/I_1)|\Omega_3|}{\sqrt{\left(1+\frac{I_3^2\Omega_3^2}{I_1^2\Omega_{\bot0}^2}\right)\exp\!\left\{\frac{4d^2I_3^2\Omega_3^2}{3I_1^3c^3}\,t\right\}-1}},
\lab{3.12}
\end{eqnarray}
where $\tilde\Omega_3\equiv(1-I_3/I_1)\Omega_3$ and $\Omega_{\bot0}=\Omega_\bot|_{t=0}$. (Here the choice of a reference frame
provides that $\Omega_1|_{t=0}>0$, $\Omega_2|_{t=0}=0$).

It follows from \re{3.11} that the precession of $\BOm_\bot$ has opposite directions for a prolate ($I_3<I_1$) and oblate ($I_3>I_1$) top,
and is absent for a spherical top.

In the limit $\Omega_3\to0$ one obtains for $\Omega_\bot$:
%
\begin{equation}\lab{3.13}
\Omega_\bot=
\frac{\Omega_{\bot0}}{\sqrt{1+\frac{4d^2\Omega_{\bot0\rule[-0.2ex]{0pt}{0pt}}^2}{3I_1c^3}\,t}}.
\end{equation}
Besides, this expression is true in the case $I_3\to0$ of spinning rod when the variable $\Omega_3$
and a direction of the vector $\BOm_\bot$ (but not its length $\Omega_\bot$) become indefinite, and eqs. \re{3.11} become meaningless.


\section{Dynamics of an axially symmetric spinning top with the transverse dipole}
\renewcommand{\theequation}{4.\arabic{equation}}
\setcounter{equation}{0}

The case of the spinning top with the dipole moment perpendicular to a symmetry axis,
%
\begin{equation}\lab{4.1}
I_{ij}=I_i\delta_{ij}, \quad I_2=I_1;\qquad d_1\equiv d,\quad d_2=d_3=0,
\end{equation}
is more cumbersome than that with parallel moment. The reduced equation of motion splits into the following ones:
%
\begin{eqnarray}
I_1\dot\Omega_1&=&(I_1-I_3)\Omega_2\Omega_3,
\lab{4.2}\\
I_1\dot\Omega_2&=&(I_3-I_1)\Omega_1\Omega_3-\frac{2d^2}{3c^3}\left\{\Omega^2+\frac{(I_1-I_3)(2I_1-I_3)}{I_1^2}\Omega_3^2\right\}\Omega_2,
\lab{4.3}\\
I_3\dot\Omega_3&=&-\frac{2d^2}{3c^3}\left\{\Omega^2+\frac{I_1-I_3}{I_1}(2\Omega_1^2-\Omega_2^2)\right\}\Omega_3.
\lab{4.4}
\end{eqnarray}

One finds easily two partial solutions of the set \re{4.2}-\re{4.4}.

1). $\Omega_1=\Omega_2=0$. Eqs. \re{4.2}, \re{4.3} become identities while \re{4.4} reduces to the equation
%
\begin{equation}\lab{4.5}
\dot\Omega_3=-\frac{2d^2}{3I_3c^3}\Omega_3^3
\end{equation}
which, upon redefinition of parameters, coincides with the flat rotator equation \re{2.3}. The power-law solution
of the type \re{2.4} tends to zero, $\Omega_3\to0$, at $t\to\infty$.

2). $\Omega_3=0$. Then $\Omega_1=\,$const by \re{4.2} while \re{4.4} reduces to the equation
%
\begin{equation}\lab{4.6}
\dot\Omega_2=-\frac{2d^2}{3I_1c^3}\{\Omega_1^2+\Omega_2^2\}\Omega_2.
\end{equation}
It possesses solution on the type \re{3.12} with the exponential asymptotics $\Omega_2\to0$.

Asymptotics of both solutions lead to a set of fixed points
$\BOm_\infty=\{\Omega_\infty, 0,0 \}$, $\Omega_\infty\in{\Bbb R}$. There are no other points which are
fixed for the set \re{4.2}-\re{4.4}.

General exact solution of the set \re{4.2}-\re{4.4} is unknown. Farther some qualitative analysis
is undertaken. For this purpose let us introduce
the dimensionless quantities:
%
\begin{equation}\lab{4.7}
\tau_0=\frac{2d^2}{3I_1c^3},\qquad
\tau=\frac{t}{\tau_0},\qquad \omega=\tau_0\Omega,
\end{equation}
and change Cartesian components $\omega_1$, $\omega_2$ by
cylindrical ones $\omega_\bot$, $\varphi$:
%
\begin{equation}\lab{4.8}
\omega_1=\omega_\bot\cos\varphi, \qquad
\omega_2=\omega_\bot\sin\varphi;
\end{equation}
here the angle $\varphi$ determines a direction of the vector
$\B\omega_\bot=\{\omega_1,\omega_2,0\}$, and $\omega_\bot=|\B\omega_\bot|$.
In these terms the equations \re{4.2}-\re{4.4} take the form:
%
\begin{eqnarray}
\dot\omega_\bot&=&-\{\omega_\bot^2+(1+\delta+\delta^2)\omega_3^2\}\omega_\bot\sin^2\!\varphi,
\lab{4.9}\\
\dot\omega_3&=&-\frac1{1-\delta}\{(1-\delta+3\delta\cos^2\!\varphi)\omega_\bot^2+\omega_3^2\}\omega_3,
\lab{4.10}\\
\dot\varphi&=&-\delta\omega_3-\frac12\{\omega_\bot^2+(1+\delta+\delta^2)\omega_3^2\}\sin2\varphi,
\lab{4.11}\\
\mbox{where}\qquad\qquad&&\delta\equiv1-I_3/I_1,\qquad -1\le\delta<1;
\lab{4.12}
\end{eqnarray}
$\delta\gtrless0$ corresponds to the prolate (oblate) inertia spheroid which becomes
the sphere at $\delta=0$ and the disc at $\delta=-1$. The case $\delta=1$ of thin rod has no meaning since
the variables $\omega_3$, $\varphi$ and the direction of dipole are indefinite.

Note some general properties of the equations \re{4.9}--\re{4.11}.

It follows from eq. \re{4.10} that $\D{}{\omega_3}/\D{}{\tau}\lesseqqgtr0$ provided $\omega_3\gtreqqless0$, i.e.,
$|\omega_3|\to0$ monotonously at $\tau\to\infty$.

It follows from eq. \re{4.9} that $\omega_\bot\to0$ monotonously at $\tau\to\infty$ with undetermined
final angle $\varphi_\infty$. Otherwise one may occur $\omega_\bot\to|\omega_\infty|>0$ provided $\sin\varphi\to0$.
But actual behavior of the angle $\varphi$ is not evident from the equation \re{4.11} and needs some analysis.

\subsection{The averaged dynamics and the linearized dynamics}

It is noteworthy that the r.-h.s of the set \re{4.9}-\re{4.11} is a
$\pi$-periodic function of $\varphi$. Averaging this set yields the
equations for averaged variables $u\equiv{\bar\omega}_\bot^2$,
$v\equiv{\bar\omega}_3^2$,   $\bar\varphi$:
%
\begin{eqnarray}
\dot u&=&-\{u+(1+\delta+\delta^2)v\}u,
\lab{4.13}\\
\dot v&=&-\frac1{1-\delta}\{(2+\delta)u+2v\}v,
\lab{4.14}\\
\dot{\bar\varphi}&=&-\delta\bar\omega_3.
\lab{4.15}
\end{eqnarray}

This closed set of equations possesses
the exact solution in a parametric form:
%
\begin{eqnarray}
u&=&\xi v = C\xi^{1-q}|(1+2\delta)\xi+1+\delta^3|^p,
\lab{4.16}\\
\tau-\tau_0&=&\frac{1-\delta}C\left|\int_{\xi_0}^\xi\D{}\xi\,\xi^{q-1}\{(1+2\delta)\xi+1+\delta^3\}^{-p-1}\right|,
\lab{4.17}\\
\bar\varphi-\bar\varphi_0&=&\pm\frac{\delta(1-\delta)}{\sqrt{C}}\int_{\xi_0}^\xi\D{}\xi\,\xi^{\frac{q}2-1}|(1+2\delta)\xi+1+\delta^3|^{-\frac{p}2-1},
\lab{4.18}\\
\mbox{where}&&q=\frac2{1+\delta^3},\quad p=\frac{\delta(1-\delta)(3+3\delta+\delta^3)}{(1+2\delta)(1+\delta)},
\lab{4.19}
\end{eqnarray}
and $C>0$ is an integration constant. Quadratures \re{4.17}--\re{4.18} can be expressed in terms of incomplete beta-functions,
but these expressions will not be needed further.

Instead, let us consider asymptotics of the solution \re{4.16}-\re{4.18} at $\tau\to\infty$:
%
\begin{eqnarray}
-1\le\delta\le-1/2)&\qquad&\bar\omega_\bot\sim|\bar\omega_3|\sim\tau^{-1/2};\quad|\bar\varphi|\sim\tau^{1/2};
\lab{4.20}\\
-1/2<\delta<1)&\qquad&\bar\omega_\bot\sim\tau^{-1/2},\quad |\bar\omega_3|\sim\tau^{-\frac{1+\delta/2}{1-\delta}},\quad |\bar\varphi|\sim\tau^{-\frac{3\delta}{2(1-\delta)}}.
\lab{4.21}
\end{eqnarray}
It is seen from \re{4.21} that the set \re{4.13}--\re{4.15} makes a sense for $\delta<0$ since $|\bar\varphi|$
is an infinitely increasing function at $\tau\to\infty$. On the contrary,
$|\bar\varphi-\bar\varphi_0|$ is asymptotically decreasing function at $\delta>0$ in which case the averaging of the set
\re{4.9}-\re{4.11} is meaningless. Nevertheless, one can expect that $\varphi$ itself tends to zero or some finite value
$\varphi_0$ corresponding to a fixed point of the original set of equations \re{4.9}-\re{4.12}.
\bigskip

Thus let us return to the equations \re{4.9}-\re{4.11} and analyze them in the neighborhood of the fixed point
$\BOm_\infty=\{\Omega_\infty, 0,0 \}$. In terms of dimensionless polar coordinates this point is determined by
$\omega_\bot=\omega_\infty\equiv\tau_0|\Omega_\infty|$, $\omega_3=0$, $\varphi=k\pi$ with $k=0,\pm1,\pm2,\dots$.
The linearization of \re{4.5}-\re{4.9} in a neighborhood of fixed points yields the set of equations
%
\begin{eqnarray}
\dot\nu&=&0,
\lab{4.22}\\
\dot\omega_3&=&-\frac{1+2\delta}{1-\delta}\omega_\infty^2\,\omega_3,
\lab{4.23}\\
\dot\psi&=&-\delta\,\omega_3-\omega_\infty^2\,\psi
\lab{4.24}
\end{eqnarray}
for the deviations $\nu=\omega_\bot-\omega_\infty$, $\omega_3=\omega_3-0$, $\psi=\varphi-k\pi$.
This set possesses exponential solutions $\propto\mathrm{e}^{\lambda\tau}$ with
the characteristics $\lambda_1=0$, $\lambda_2=-\frac{1+2\delta}{1-\delta}\omega_\infty^2$,
$\lambda_3=-\omega_\infty^2$. The root $\lambda_1=0$ generates the change of the fixed value
$\omega_\infty\to\omega_\infty+\nu_0$ which corresponds to a nearby fixed point. Fixed points with $\omega_\infty\ne0$ are stable
(due to the linear approximation) for $\delta>-1/2$ and unstable for $\delta\le-1/2$.
Once the system gets into the neighborhood of a stable fixed point, the latter will be reached necessarily.

\subsection{Analytical vs numerical results}

The analysis of the original \re{4.9}--\re{4.11}, the averaged \re{4.13}--\re{4.15} and
the linearized \re{4.22}--\re{4.24} sets of equations shows that there are tree different cases.
1).~If $0\le\delta<1$, i.e., the spinning top is prolate or spherical ($0<I_3\le I_1$),
the fixed point $\BOm_\infty=\{\Omega_\infty, 0,0 \}$ with any $\Omega_\infty\ne0$ is stable.
Thus, starting from arbitrary $\BOm_0=\{\Omega_{10},\Omega_{20},\Omega_{30}\}$, after few revolutions,
the vector $\BOm$ tends to $\BOm_\infty$ exponentially with the characteristic braking time $T=\{(3I_1/I_3-2)\tau_0\Omega_\infty^2\}^{-1}$.
It is noteworthy that the asymptotic components arises $\Omega_1|_{t\to\infty}\equiv\Omega_\infty\ne0$ even if $\Omega_{10}=0$.
2).~For the markedly oblate top of $-1\le\delta\le-1/2$ (i.e., $\frac32I_1\le I_3\le2I_1$)
this fixed point $\BOm_\infty=\{\Omega_\infty, 0,0 \}$ with $\Omega_\infty\ne0$
is unstable. Thus, by general properties of the set \re{4.9}--\re{4.11} and of \re{4.16}--\re{4.18},
the vector $\BOm$ continues to decrease by the asymptotic power law $\Omega\sim1/\sqrt{\tau_0t}$ and reaches the value
$\B\Omega=0$ after an infinite number of revolutions around the axis O$\rho_3$.
3).~In the case $-1/2<\delta<0$ of a weakly oblate top (i.e., $I_1<I_3<\frac32I_1$) both scenarios are possible,
and the final state depends on the starting point.
%
%
These cases are summarized in figure \ref{fig1}.
%
\begin{figure}[b]
\begin{center}
\includegraphics[scale=0.5]{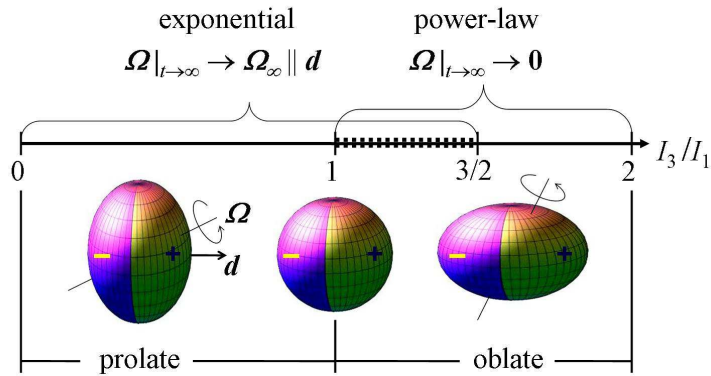}
\caption{Asymptotic behavior of an axially symmetric top with the transverse dipole moment $\B d$ depending
on the aspect ratio $I_3/I_1$ of the inertia spheroid.
}\lab{fig1}
\end{center}
\end{figure}
%
\begin{figure}[ht]
\begin{center}
\includegraphics[scale=1.1]{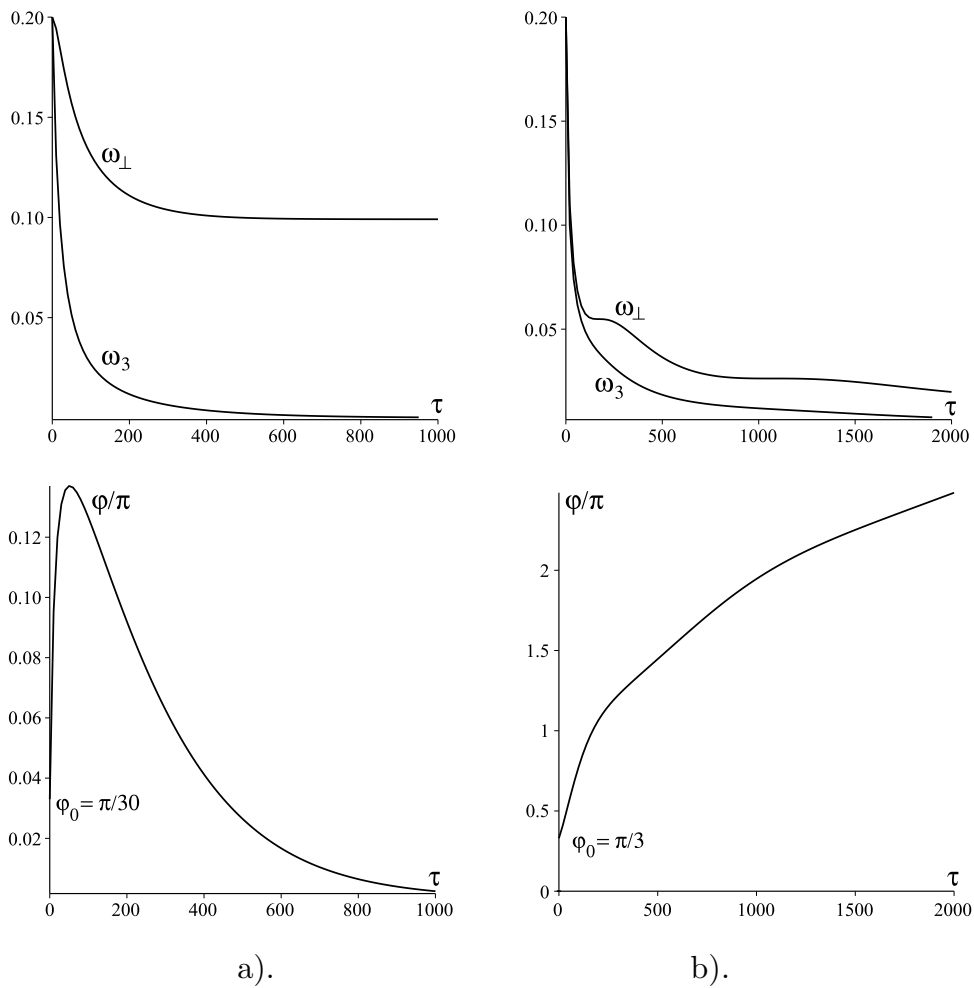}
\caption{Evolution of the axially symmetric top with the transverse dipole moment for $\delta=-1/5$,
$\omega_{\bot\,0}=\omega_{3\,0}=1/5$. a). $\varphi_0=\pi/30$.  b). $\varphi_0=\pi/3$.
}\lab{fig2}
\end{center}
\end{figure}
Numerical integration of the set \re{4.9}--\re{4.11} with initial conditions chosen
randomly confirms this picture. There are shown at the figure \ref{fig2} two examples of evolution of the top with
$\delta=-1/5$ for the same dimensionless initial values $\omega_{\bot\,0}=\omega_{3\,0}=1/5$ but with different
$\varphi_0=\pi/30$ and $\varphi_0=\pi/3$. The example a) is typical for the prolate top, the b) -- for the markedly oblate top.


\section{Conclusions}

The equation of a rotary motion of the rigid body with the electric dipole moment \re{3.3}
is derived from the Landau-Lifshitz angular momentum balance condition \cite{L-L87E}.
This equation of the 2nd order with respect to the angular velocity $\BOm$ can be reduced approximately
to the 1st-order Euler-type equation.

The specific case of a body with the axially-symmetric inertia ellipsoid is considered.
The Euler equations for the spinning top with the longitudinal dipole is axially-symmetric
and integrable. The longitudinal component $\Omega_3$ of the angular velocity $\BOm$ is conserved
while the transverse component $\Omega_\bot$ decreases exponentially at $t\to\infty$ with the braking time
$T=1/(\tau_0\Omega_3^2I_3^2/I_1^2)$, where $\tau_0$ is defined by \re{4.7}.
If $\Omega_3=0$, then $\Omega$ decreases asymptotically (at $t\ll T$) by the power law $\Omega\sim1/\sqrt{\tau_0t}$
independently of the initial value of $\Omega$, similarly to the case of a plane rotator.

Absence of a symmetry may complicate the dynamics.
Here it is considered the example of an axially-symmetric top but with transverse dipole which breaks the symmetry.
In this case the dynamics is not integrable
and has been analyzed qualitatively and numerically. It turned out that a final state of the top
depends notably on its shape. If the spinning top is prolate, i.e., $I_3<I_1$, some residual component of the
the angular velocity $\BOm$ along the dipole survives (even if this component was absent initially) while other components decrease exponentially.
The markedly oblate top (i.e., if $I_3>\frac32I_1$) stops asymptotically by the power law $\Omega\sim1/\sqrt{\tau_0t}$.
For a weakly oblate top (i.e., if $I_1<I_3<\frac32I_1$) both scenarios are possible, and the final state depends
the initial conditions.

The most realistic case is the asymmetric top. Its dynamics is more complicated.
Preliminary calculations reveal exponential drift to some residual rotation if
the top is polarized along a stable principal axis. Otherwise, the top slows down
to a complete stop. Study of this braking in detail would be desirable.

Finally, let us consider some numerical estimates relevant to the experiment \cite{RDHD18} with the silica particle mentioned in the Introduction.
The inertia moment of a spherical particle is $I=\frac25mR^2$, where $m=1\,$fg and $R=50\,$nm.
Being ionized once up to the elementary charge $q=e=4.8\cdot10^{-10}\,$CGSE, the particle acquires the dipole moment $d=eR\approx2400\,$D.
Then the scale time $\tau_0\approx10^{-35}\,$s is extremely small, but the characteristic braking time for $\Omega_0=2\pi\,$GHz is astronomical:
$T\approx2\cdot10^{15}\,\rm{s}\sim0.5\cdot10^8\,$years. Hypothetically, these figures may have relevance to the interstellar dust
which consists of silicate-graphite grains of size 50-500 nm \cite{Dra03}. Grains can be ionized by cosmic rays \cite{IPGC15}
and spinned up by circularly polarized radiation from relativistic sources such as a black hole \cite{BFB99}.
But GHz rotation of grains seems unlikely.

More realistic is a laboratory spin up (and subsequent slowdown) of
artificial nanoparticles which may carry a large permanent electric
dipole moment. The examples are Janus-like particles \cite{L-H11} or
nanocrystals CdSe and CdS which at the size $\sim\,$5\,nm may
acquire up to 100\,D of a dipole moment \cite{S-K06}. Of special
interest are organic nanocrystals. Cellulose
$\sim\,$300\,nm$\times$30\,nm--elongated nanocrystals are reported
to possess the moment $\gtrsim$\,4000\,D \cite{F-PBL14}. But
DAST-nanoparticles may appear the record holders: such a
100\,nm$\times$100\,nm$\times$50\,nm--crystal is estimated to have
the moment $\gtrsim\,2.8\cdot10^7$\,D \cite[pp.\,387-390]{MNS03}.
The characteristic braking time for this last
example with the initial $\Omega_0=2\pi\,$GHz is $T\sim\,0.3\,$year
which is of order of a storage time in Penning trap \cite{Haf03}.
Since the idea of a similar trap for neutral dipole particles is
developing at present \cite{PMY2019}, the rotational braking of such
particles by radiation reaction can be of interest.


\section*{Appendix}

\subsection*{A. Analysis of the plane rotator Lorentz equation}
\renewcommand{\theequation}{A.\arabic{equation}}
\setcounter{equation}{0}

In terms of dimensionless variables $\tau=t/\tau_0$, $\omega=\tau_0\Omega$ the
nonlinear differential 2nd-order equation \re{2.2} becomes free of any parameter:
%
\begin{equation}\lab{A.1}
\ddot\omega-\dot\omega-\omega^3=0;
\end{equation}
here the dot ``$\quad\dot{}\quad$'' denotes differentiation  by $\tau$.
The equation \re{A.1} is invariant under time translations $\tau\to\tau+\lambda$,
$\lambda\in\Bbb R$ but the corresponding integral of motion is unknown.
The change of variable $\tau\to\theta={\mathrm e}^\tau$ reduces the equation \re{A.1}
to the form:
%
\begin{equation}\lab{A.2}
\D{2}{\omega}/\D{}{\theta^2}=\omega^3/\theta^2,
\end{equation}
which is of the Emden-Fowler type equation $\D{2}{y}/\D{}{x^2}=x^ny^m$. The pair of indices $n{=}-2$,
$m=3$ does not correspond to integrable cases of this equation \cite{P-Z03}.

Let us study the asymptotic behavior of solutions \cite{Bel53} of
the equations \re{2.2} or \re{A.1} at
$t\to\pm\infty,0$.
We suppose a power-law or exponential asymptotic behavior of solutions.
\smallskip

\noindent
$\boldsymbol{t\to+\infty}$). Substituting the power-law anzatz
$\omega=A\,\tau^\alpha[1+O(\tau^{-1})]$
with the constants $A$ and $\alpha$ to be found into eq. \re{A.1} leads to the equality:
%
\begin{eqnarray}\lab{A.3}
 A\alpha(\alpha-1)\tau^{\alpha-2} - A\alpha\tau^{\alpha-1}
- A^3\tau^{3\alpha} =O(\tau^{\alpha-3}) + O(\tau^{\alpha-2}) +
O(\tau^{3\alpha-1}).\hspace{5ex}
\end{eqnarray}
The 1st term in l.-h.s. is negligibly small as to the 2nd term which,
in turn, can be canceled by the 3rd term, provided $\alpha=-1/2$ and $A^2=1/2$.
In dimensional terms this can be summarized as follows:
%
\begin{equation}\lab{A.4}
\dot\Omega\sim-\tau_0\Omega^3 \ \Rightarrow\ \Omega(t)
=\pm\frac{1}{\sqrt{2\tau_0t}}[1+O(t^{-1})],\quad t\to+\infty.
\end{equation}
%
%
$\boldsymbol{t\to0}$). Similarly, one obtains the asymptotics:
$\displaystyle
\omega(\tau)=\pm\frac{\sqrt{2}}{\tau}[1+O(\tau)]
$.\\
Since the equation \re{2.2} is invariant under the time translation
$t\to t-t_1$ by the arbitrary $t_1$, this asymptotics can be presented in dimensional terms
as follows:
%
\begin{equation}\lab{A.5}
\ddot\Omega\sim\Omega^3 \ \Rightarrow\
\Omega(t)=\pm\frac{\sqrt{2}}{t-t_1}[1+O(t{-}t_1)],\quad t\to t_1,\
\forall t_1\in{\Bbb R}.
\end{equation}
%
$\boldsymbol{t\to-\infty}$).
The equation \re{2.2} or \re{A.1} does not admit a
power-law asymptotics, but the Emden-Fowler equation
\re{A.2} does:
$
\omega\sim A\theta=A{\mathrm e}^\tau$, $\theta\to+0 \
\Leftrightarrow\ \tau\to-\infty
$,
where $A$ is an arbitrary real constant. In dimensional terms we have:
%
\begin{equation}\lab{A.6}
\dot\Omega\sim\tau_0\ddot\Omega \ \Rightarrow\ \Omega(t)\sim
A\exp(t/\tau_0)/\tau_0,\quad t\to-\infty.
\end{equation}

The asymptotics \re{A.5} describes unlimited self-acceleration
of a circulating particle during a finite time.
Remarkably, the small parameter
$\tau_0$ drops out from this asymptotics. Thus the corresponding solution
must be regarded as obviously nonphysical.

The asymptotics \re{A.6} is obviously non-analytical in $\tau_0$. It is a segment of
non-physical solution of eq. \re{2.2} which, in turn, is similar
(at $ t\to-\infty$) to the runaway solution of the original Lorentz equation \re{2.1}.

The only asymptotics \re{A.4} is physically meaningful since it correlates completely
with the solution \re{2.4} of the reduced equation \re{2.3}.

More details on solutions of the equation \re{A.1} can be seen in the phase portrait
of the rotator. Let us recast for this purpose the 2nd order (with respect to $\omega$)
equation \re{A.1} into the dynamical system:
%
\begin{eqnarray}
\dot\omega&=&\varpi,
\lab{A.7}\\
\dot\varpi&=&\varpi+\omega^3. \lab{A.8}
\end{eqnarray}
It follows from this the Abel equation for phase trajectories of the system:
%
\begin{equation}\lab{A.9}
\frac{\D{}{\varpi}}{\D{}{\omega}}=1+\frac{\omega^3}{\varpi}.
\end{equation}
By now this equation is not solvable, but the asymptotics
\re{A.4}, \re{A.5}, \re{A.6} suggest corresponding
asymptotics of phase trajectories:
%
\begin{eqnarray}
\tau\to+\infty)&\quad&\left.{\omega\sim\pm(2\tau)^{-1/2}\to0,\atop
\varpi\sim\mp(2\tau)^{-3/2}\to0}\right\}\quad\Rightarrow\quad
\varpi\sim-\omega^3.\hspace{5ex} \lab{A.10}\\
\forall\tau_1\quad\bar\tau\equiv\tau{-}\tau_1\to0)&\quad&\left.
{\omega\sim\pm\sqrt{2}/\bar\tau\to\pm\infty,\atop
\varpi\sim\mp\sqrt{2}/{\bar\tau}^2\to\mp\infty}\right\}\quad\Rightarrow\quad
\varpi\sim\mp\frac{\omega^2}{\sqrt{2}}.
\lab{A.11}\\
\tau\to-\infty)&\quad&\left.
{\omega\sim A{\mathrm e}^\tau\to0,\atop \varpi\sim A{\mathrm
e}^\tau\to0}\right\}\qquad\quad\ \Longrightarrow\quad \varpi\sim\omega.
\lab{A.12}
\end{eqnarray}

The system possesses a fixed point $O=(\omega\,{=}\,0, \varpi\,{=}\,0)$, which is unstable.
This follows from the behavior in neighborhood of this point of the following
Lyapunov function \cite{A-P82}:
%
\begin{eqnarray}
&&V(\omega,\varpi)=\omega(\varpi-\ha\omega):
\lab{A.13}\\
&&V(0,0)=0; \quad V(0\lessgtr\omega\lessgtr2\varpi)>0;\quad \dot
V=\varpi^2+\omega^4>0. \lab{A.14}
\end{eqnarray}

A phase portrait of a charged plane rotator derived by means of
the above analysis and a
numerical integration of the equation \re{A.9} is presented in
figure \ref{fig3}. It is divided by four domains by two separatrices
$AOC$ and $BOD$ crossing in the fixed point $O$.
In the neighborhood of $O$ the separatrix $AOC$ is described by
the asymptotics \re{A.10} while $BOD$ by the asymptotics \re{A.12}.
Opposite i.e. infinite asymptotics of these separatrices as well as
the asymptotics of other phase trejectories are described by eq.
\re{A.11}, and are reachable in a finite time. Thus all the phase
trajectories are non-physical, except the separatrix $AOC$.
%
\begin{figure}
\begin{center}
\includegraphics[scale=0.53]{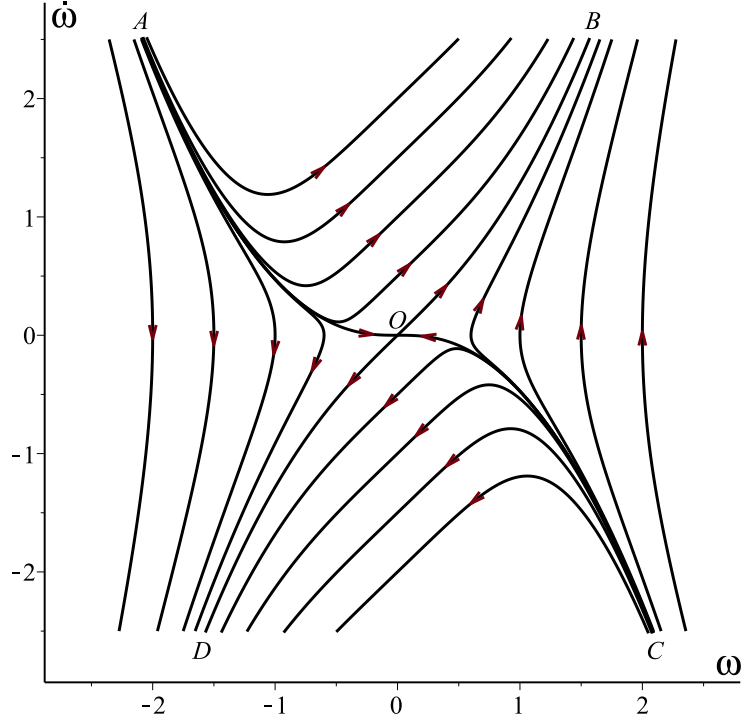}
\vspace{-2ex} \caption{Phase portrait of the charged plane
rotator.\vspace{-3ex}} \lab{fig3}
\end{center}
\end{figure}

It is obvious from figure \ref{fig3} that the set (curve) $AOC$ is a
repeller, or, following the Spoon's terminology, a critical manifold
\cite{Spo00}. Every solution passing through an arbitrary point
beyond the curve $AOC$ goes to infinity in a finite time, so it is
non-physical.

Every physical solution passes points of the curve $AOC$.
Thus it satisfies both the 2st-order equation \re{A.1}
as well as the 1st-order equation

\begin{equation}\lab{A.15}
\dot\omega=f(\omega),
\end{equation}
where the function $f(\omega)$ determines the curve $AOC$
by the equation $\varpi=f(\omega)$ and thus
meets the following conditions:

%
\begin{eqnarray}
&&f(\omega)=-\omega^3+f(\omega)\,\D{}{\!f(\omega)}\!/\D{}{\omega},
\lab{A.16}\\
&&f(\omega)
\mathop{\smash{{\rule{2ex}{0pt}\atop\widetilde{\rule{0ex}{1.5ex}\phantom{==}}}}}\limits_{\omega\to0}
-\omega^3.
\lab{A.17}
\end{eqnarray}
In other words, the equation \re{A.15}-\re{A.17} is the exact 1st-order reduction
of the 2nd-order equation \re{A.1}.
Since the equation \re{A.16} or \re{A.9} is not integrable, one can
use an analytic approximation for $f(\omega)$. The approximation $f(\omega)\approx-\omega^3$
represents in eq. \re{A.15} a dimensionless form of the reduced equation \re{2.2}.
It is precise in the neighborhood of $\omega\to0$ and yields the asymptotics
\re{A.10}. Both asymptotics \re{A.10} and \re{A.11} can be taken into account
in the following approximation $f(\omega)\approx-\omega^3/\sqrt{1+2\omega^2}$
with which the equation \re{A.15} is integrable analytically.

\subsection*{B. Analysis of the Euler-Lorentz equations for a longitudinal dipole}
\renewcommand{\theequation}{B.\arabic{equation}}
\setcounter{equation}{0}

The set of equations \re{3.6}--\re{3.8} is not solved exactly.
Here an asymptotic behavior of solutions
at $t\to\pm\infty,0$ will be analyzed.
In terms of dimensionless cylindrical variables \re{4.7}, \re{4.8}
the equation \re{3.6}--\re{3.8} take the form:
%
\begin{eqnarray}
&&\ddot\omega_\bot-\dot\omega_\bot-\omega_\bot^3-\omega_\bot\nu^2=0,
\lab{B.1}\\
&&\omega_\bot(\dot\nu-\nu+\zeta\omega_3)+2\dot\omega_\bot\nu=0,
\lab{B.2}\\
&&\zeta\dot\omega_3=0,
\lab{B.3}
\end{eqnarray}
where $\nu\equiv\dot\varphi+\omega_3$, $\zeta=I_3/I_1$, and the dot ``$\hspace{1ex}\dot{}\hspace{1ex}$'' denotes the differentiation over $\tau$.
\smallskip

\noindent $\boldsymbol{\tau\to+\infty}$). The set
\re{B.1}--\re{B.2} admits a power-law asymptotics provided
$\omega_3=0$ and/or $\zeta=0$. In these cases the equation \re{B.2}
possesses the integral of motion $C=\omega_\bot^2\nu{\mathrm e}^{-\tau}$
which permits us to eliminate $\nu$ from the equation \re{B.1}:
%
\begin{equation}\lab{B.4}
\ddot\omega_\bot-\dot\omega_\bot-\omega_\bot^3-C^2{\mathrm e}^{2\tau}/\omega_\bot^3=0.
\end{equation}
This equation admits a power-law asymptotics at the only value
$C=0$. Then it becomes identical to the equation \re{A.1} for
a plane rotator. Thus we have:
%
\begin{eqnarray}
&&\omega_\bot=\frac{\pm1}{\sqrt{2\tau}}[1+O(\tau^{-1})],
\lab{B.5}\\
&&\varphi=\varphi_0\qquad \mbox{if}\quad \omega_3=0,\quad\zeta\ne0.
\lab{B.6}
\end{eqnarray}
At $\zeta=0$ the solution $\varphi=\varphi_0 -\int_0^\tau\D{}{\tau}\omega_3(\tau)$
loses a meaning since $\varphi$ is unobservable.

In the case $\zeta\ne0$, $\omega_3=\,$const$\,\ne0$ the set \re{B.1}--\re{B.2}
does not admit a power-law asymptotics at $\tau\to+\infty$.
Thus let us look for the exponential asymptotics. The change of the variable
$\tau$ by $\theta={\mathrm e}^\tau$ reduces the equations
\re{B.1}--\re{B.2} to the form:
%
\begin{eqnarray}
&&\theta^2\omega''_\bot-\omega_\bot^3-\omega_\bot\nu^2=0,
\lab{B.7}\\
&&\omega_\bot(\theta\nu'-\nu+\zeta\omega_3)+2\theta\omega'_\bot\nu=0,
\lab{B.8}
\end{eqnarray}
where the prime ``$\ \ {}'\ \ $'' denotes the differentiation with respect
to $\theta$. For asympotices (at $\theta\to\infty$, i.e., $\tau\to\infty$) for
$\omega_\bot$ and $\nu$ we use the ansatzes:
%
\begin{equation}\lab{B.9}
\omega_\bot=A\theta^\alpha[1+O(\theta^{-1})], \qquad
\nu=B\theta^\beta[1+O(\theta^{-1})],
\end{equation}
where $A$, $\alpha$, $B$, $\beta$ are real constants to be found.
The substitution of these anzatzes into the set \re{B.7}--\re{B.8}
leads to the equations:
%
\begin{eqnarray*}
&&\alpha(\alpha-1)-A^2\theta^{2\alpha}
-B^2\theta^{2\beta}=O(\theta^{-1})+O(\theta^{2\alpha-1})+O(\theta^{2\beta-1}),
\\
&&(2\alpha+\beta-1)B\theta^\beta+\zeta\omega_3=O(\theta^{\beta-1})+O(\theta^{-1}).
\end{eqnarray*}
They are compatible at the only values $\beta=0$, $\alpha<0$, and lead to a
4th-degree equation for $\alpha$ with two real roots:
%
\begin{equation}\lab{B.10}
(2\alpha-1)^2\alpha(\alpha-1)=(\zeta\omega_3)^2 \quad \Longrightarrow \quad
\alpha_\pm=\frac12\pm\frac12\sqrt{\frac12\!\left(1+\sqrt{1+(4\zeta\omega_3)^2}\right)}\gtrless0.
\end{equation}
In the present case the only $\alpha_-$ is relevant, and we have:
%
\begin{equation}\lab{B.11}
\alpha_-\approx-(\zeta\omega_3)^2+5(\zeta\omega_3)^4+\dots; \
B=\frac{\zeta\omega_3}{1-2\alpha_-}\approx \zeta\omega_3-2(\zeta\omega_3)^3+\dots,
\end{equation}
If a value of the variable $\omega_3=\tau_0\Omega_3$ is small, one
can retain only leading terms of these expansions, and so obtain
from \re{B.9}  the asymptotics:
%
\begin{equation}\lab{B.12}
\omega_\bot\sim A\theta^{\alpha_-}\approx A{\mathrm e}^{-(\zeta\omega_3)^2\tau}, \qquad
\varphi-\varphi_0=-(\omega_3-B)\tau\approx-(1-\zeta)\omega_3\tau,
\end{equation}
where $A$ is an arbitrary constant.
\medskip

{\em Non-physical asymptotics} for the top and the rotator are
similar. If $\zeta\ne0$:
%
\begin{eqnarray}
\B{\tau\to0)}&&\omega_\bot\sim \pm\frac{\sqrt2}{\tau}[1+O(\tau)]; \qquad \varphi-\varphi_0=-(1-\zeta)\omega_3\tau+O(\tau^2);\hspace{6ex}
\lab{B.13}\\
\lab{B.14}
\B{\tau\to-\infty)}&&\omega_\bot\sim A\theta^{\alpha_+}\approx A{\mathrm e}^\tau, \qquad\quad\
\varphi-\varphi_0\approx-(1+\zeta)\omega_3\tau,
\end{eqnarray}
where $\alpha_+$ is defined in eq. \re{B.10}.
If $\zeta=0$ expressions for $\varphi$ in eqs. \re{B.13}--\re{B.14} lose meaning.

\section*{Acknowledgements}
I would like to thank Profs. M. Przybylska, A. J. Maciejewski, A. Panasyuk, Drs. R. Matsyuk, Yu. Yaremko
and all the staff of the Department for Theoretical Physics of Ivan Franko National University of Lviv
for discussions of this work and fruitful suggestions.


\begin{thebibliography}{10}
\providecommand{\url}[1]{{#1}}
\providecommand{\urlprefix}{URL }
\expandafter\ifx\csname urlstyle\endcsname\relax
  \providecommand{\doi}[1]{DOI \discretionary{}{}{}#1}\else
  \providecommand{\doi}{DOI \discretionary{}{}{}\begingroup
  \urlstyle{rm}\Url}\fi

\bibitem{RDHD18}
R.~Reimann, M.~Doderer, E.~Hebestrait, R.~Diehl, M.~Frimmer, D.~Windey,
  F.~Tebbenjohanns, L.~Novotny, Phys. Rev. Lett. \textbf{121}(3), 033602 (2018)

\bibitem{AXBJGL20}
J.~Ahn, Z.~Xu, J.~Bang, P.~Ju, X.~Gao, T.~Li,
Nat. Nanotechnol. \textbf{15}(2), 89 (2020)

\bibitem{Roh90}
F.~Rohrlich, \emph{Classical charged particles: foundations of their theory}
  (Addison-Wesley, New York, 1990)

\bibitem{YaT12}
Y.~Yaremko, V.~Tretyak, \emph{Radiation reaction in classical field theory:
  basics, concepts, methods} (LAP, Saarbr{\"u}cken, 2012)

\bibitem{L-L87E}
L.D. Landau, E.M. Lifshitz, \emph{The clasical theory of fields}, \emph{-- {\em
  Course of theoretical physics}}, vol.~2, 4th edn. (Butterworth-Heinemann,
  1987)

\bibitem{Spo00}
H.~Spohn, Found. Phys. \textbf{36}(10), 1474 (2006)

\bibitem{Par06}
G.A. {\relax de Parga}, Europhys. Lett. \textbf{50}(3), 287 (2000)

\bibitem{dGS72}
S.R. \relax{de G}root, L.G. Suttorp, \emph{Foundations of electrodynamics}
  (North-Holland Publ. Co., Amsterdam, 1972)

\bibitem{Dra03}
B.T. Draine, Annu. Rev. Astron. Astr. \textbf{41}, 241 (2003)

\bibitem{IPGC15}
A.V. Ivlev, M.~Padovani, D.~Galli, P.~Caselli, Astrophys. J. \textbf{812}(2),
  135 (2015)

\bibitem{BFB99}
G.C. Bower, H.~Falcke, D.C. Backer, Astrophys. J. Lett. \textbf{523}(1), L29
  (1999)

\bibitem{L-H11}
M.~Lattuada, T.A. Hatton, Nano Today \textbf{6}, 286—308 (2011)

\bibitem{S-K06}
S.~Shanbhag, N.A. Kotov, Psys. Chem. Lett. B \textbf{110}, 12211 (2006)

\bibitem{F-PBL14}
B.~Frka-Petesic, B.~Jean, L.~Heux, Europhys. Lett. \textbf{107}(2), 28006
  (2014)

\bibitem{MNS03}
H.~Masuhara, H.~Nakanishi, K.~Sasaki, \emph{Single Organic Nanoparticles}
  (Springer-Verlag, Berlin, 2003)

\bibitem{Haf03}
H.~H{\"a}ffne{r\ et\ al.}, Eur. Phys. J. D \textbf{22}(2), 163 (2003)

\bibitem{PMY2019}
M.~Przybylska, A.J. Maciejewski, {\relax Yu}.~Yaremko, in \emph{12$^{th}$
  Workshop on Current Problems in Physics. Book of Abstracts} (Zielona
  G{\relax\'o}ra, Poland, 14 - 17 October 2019), p.~46.
\newblock {\relax h}ttp://www.if.uz.zgora.pl/$\sim$wcpp/wcpp19/abstracts.pdf

\bibitem{P-Z03}
A.D. Polyanin, V.F. Zaitsev, \emph{Handbook of Exact Solutions for Ordinary
  Differential Equations}, 2nd edn. (Chapman \& Hall/CRC, Boca Raton, 2003)

\bibitem{Bel53}
R.~Bellman, \emph{Stability theory of differential equations} (McGraw-Hill, New
  York, 1953)

\bibitem{A-P82}
D.K. Arrowsmith, C.M. Place, \emph{Ordinary differential equations. {A}
  qualitative approach with applications} (Chapman \& Hall, London New York,
  1982)

\end{thebibliography}


\end{document}